\documentclass[12pt]{iopart}
\usepackage{bm}

\begin{document}

\title{Hubbard-U band-structure methods}

\author{R C Albers$^1$, N E Christensen$^2$, and A Svane$^2$} 

\address{$^1$Theoretical Division, Los Alamos National Laboratory, Los Alamos, 
NM 87545, USA}

\address{$^2$Department of Physics and Astronomy, Aarhus University, Denmark}

\ead{rca@lanl.gov}


\begin{abstract}

The last decade has seen a large increase in the number of electronic-structure calculations that
involve adding a Hubbard term to the local density approximation band-structure
Hamiltonian.  The Hubbard term is then solved either at the mean-field level or with sophisticated
many-body techniques such as dynamical mean field theory.
We review the physics underlying these approaches and discuss their strengths
and weaknesses in terms of the larger issues of electronic structure that they
involve.  In particular,
we argue that the common assumptions made to justify such calculations
are inconsistent with what the calculations actually do.  Although many of these calculations are
often treated as essentially first-principles calculations, in fact,
we argue that they should be viewed from
an entirely different point of view, viz., as phenomenological many-body corrections to
band-structure theory.  Alternatively, they may also be considered
to be just a more complex Hubbard model than the simple one- or few-band models traditionally
used in many-body theories of solids.

\end{abstract}

\pacs{71.10.-w,71.15.-m,71.27.+a,71.28.+d}

\submitto{\JPCM}

\maketitle


\section{Introduction}

Since the very early days of quantum mechanics, the exact electronic Hamiltonian, written in terms of
the kinetic energy and electrostatic interactions between the electrons and with the nucleus,
has been known.  
In non-relativistic second-quantized form, for example, this can be written as
\begin{eqnarray}
\hat H &=& \sum_{\sigma} \int d\bm{r} \psi^{\dagger}_\sigma (\bm{r})
[ \frac{-\hbar^2}{2m} \nabla^2 + V_N(\bm{r}) - \mu] \psi_\sigma (\bm{r}) \nonumber \\
 &+& \frac{1}{2} \sum_{\sigma \sigma^\prime} \int d\bm{r} d\bm{r^\prime}
\psi^{\dagger}_\sigma (\bm{r}) \psi^{\dagger}_{\sigma^\prime} (\bm{r^\prime})
V_C (\bm{r}-\bm{r^\prime}) 
\psi_{\sigma^\prime} (\bm{r^\prime}) \psi_\sigma (\bm{r}),
\label{exact-eqn}
\end{eqnarray}
where $V_N(\bm{r})$ is the Coulomb interaction between the electrons and the nuclei,
$V_C (\bm{r}-\bm{r^\prime})$ is the Coulomb interaction ($e^2/|\bm{r}-\bm{r^\prime}|$) between the electrons,
$\psi^{\dagger}_\sigma (\bm{r})$ and $\psi_\sigma (\bm{r})$ are
creation and destruction operators for an electron at $\bm{r}$ with spin $\sigma$,
and $\mu$ is the chemical potential.
Since this Hamiltonian has strong interactions between all the electrons
in a solid, it is fundamentally a many-body problem by nature, and is too intractable to
be capable of exact solution.  Density functional theory and its application to calculations
of the electronic band structure (BS) of materials\cite{hohenberg-kohn,kohn-sham,jones89}
has had a profound impact on Condensed Matter Physics.
With fast computers, excellent numerical algorithms, and good basis sets for expanding the resulting
one-particle equations, it has been possible to reliably predict many properties of materials
from first-principles, such as
the correct ground-state crystal structure
and good values for the internal atomic positions and lattice
parameters by minimizing the total energy of the band structure.

Nonetheless, despite the excellent success of the
local density approximation (LDA) band-structure (LDA-BS) theory
in this respect (note that by the term LDA we also include
the gradient corrected or GGA versions of this theory), there have been many well
known examples of its failures as well, such as band gaps in
semiconductors that are about 50--80\% smaller than the experimentally measured
values and an incorrect description of many aspects of the electronic
structure of strongly correlated electron systems such as mixed-valence, transition-metal
oxides, heavy fermions,
and high-temperature superconductors.  These failures can usually
either be attributed to the use of a local potential to represent exchange,
which is actually non-local, or more importantly
to not adequately treating the
many-body electronic correlations.  Most involve experiments aimed at
some type of spectroscopic or excited-state properties of the electronic
quasiparticles such as photoemission or the presence or absence of band gaps.

To remedy these problems, many new techniques have been developed, particularly
with respect to strongly correlated
electron systems. These methods usually add a model Hubbard-Hamiltonian
term to the band-structure Hamiltonian, and hence have the generic form:
\begin{eqnarray}
\hat H &=& \sum_{\bm{k},ilm_l\sigma,i'l'm_{l'}\sigma'} 
[H^0(\bm{k})]_{ilm_l\sigma,i'l'm_{l'}\sigma'}
\hat c_{\bm{k}ilm_l\sigma}^\dagger \hat c_{\bm{k}i'l'm_{l'}\sigma'} \nonumber \\
&+& \frac{1}{2} \sum_{i,m\sigma\ne m'\sigma'} U_{i m\sigma,i m'\sigma'}
\hat n_{im\sigma} \hat n_{im'\sigma'}
+ \hat V_{DC}.
\label{HU-BS-eqn}
\end{eqnarray}
The first term is an LDA one-electron band-structure Hamiltonian
summed over Bloch vectors $\bm{k}$
and with orbitals at lattice sites $i$, orbital momentum $l$,
azimuthal quantum number $m_l$, and spin $\sigma$. 
The second term is a Hubbard-$U$ term that only applies to the $f$ orbitals, which we will
use as a prototype for the strongly correlated orbitals (in many of the materials
mentioned above, these orbitals are actually $d$ orbitals).
To simplify the discussion we use an orthogonalized form of the Hamiltonian
that ignores any potential overlap integrals between the orbitals.
The term $V_{DC}$ is usually called the ``double-counting" correction term
(for a fairly complete discussion of this term in the literature, 
see Ref.~\cite{petukhov03}, and references therein).
We will refer to this Hamiltonian as a Hubbard-$U$ band-structure (HU-BS) Hamiltonian.
Different methods solve this Hubbard-$U$ term to various degrees of sophistication.

The techniques involving the HU-BS Hamiltonian have been considered by many to be a
revolutionary new devlopment in electronic structure theory, especially for strongly
correlated electronic systems.  A recent review article by Held on dynamical
mean-field theory (DMFT)\cite{held07}, for example, ends with a claim that 
is commonly echoed in many places, viz., that ``the advances in electronic
structure calculations through DMFT put our ability to predict physical
quantities of such strongly correlated materials onto a similar level
as conventional electronic structure calculations for weakly correlated
materials---at last."

Given such optimism in the field, we feel that it is timely to
review these new HU-BS methods from the perspective of basic
electronic structure theory.  Because the mathematical details of the methods
have already been heavily reviewed many times recently, we will
focus this review primarily on more fundamental aspects.  
Given the history of electronic
structure methods and what we know about the underlying theory, what is
the role and usefullness of HU-BS approaches?  How predictive are they
in practice?  How much can we trust
the results of such theories and how optimistic can we be that they
represent the revolutionary breakthrough ascribed to them?  In addition, how
seriously should we view the development of even more sophisticated
methods based on this approach, especially in light of considerations
involving the underlying
foundations of the starting Hamiltonian upon which these sophisticated mathematical
tools are employed?

Finally, to be clear about the focus of this review,  we note that
we will ignore electron-phonon and other vibrational aspects of electronic structure in
this paper, as well as pairing and superconductivity, and will only consider
the case where all of the atoms are at static positions within the unit cell
of a periodic solid.

\section{Failures of Band Structure}

Before turning to specific aspects of the HU-BS methods, it is useful to begin by reviewing
the failures of conventional LDA-BS methods that motivate the search
for improvements.  By understanding what has gone wrong, we will gain insight
into what the HU-BS methods are attempting to achieve.

A brief catalog of typical failures include: (1) BS predictions of metallic materials that are
experimentally known to be insulating (e.g., CoO and FeO),
(2) absence of magnetism for materials that are magnetic (e.g., for many undoped
high-temperature superconducting oxides) and vice versa (such as Pu),
(3) band gaps that are much too small compared with experiment (e.g., for many semiconductors),
(4) electronic specific heats that are drastically too small (e.g., for heavy fermion materials),
(5) missing peaks at the Fermi energy (e.g., Kondo-like peaks),
and (6) missing satellite spectra (e.g., as occurs in Ni).
More examples can no doubt be found, but this list suffices for our purposes.

When examining this list, it becomes clear that many of the problems listed have to do
with the spectral properties of the electronic structure of a material.
However, as explained carefully in the early classic papers\cite{hohenberg-kohn,kohn-sham}
on LDA, this type of theory is designed to minimize the total ground-state
energy of the electrons in a material as a functional
of the spatial distribution of the number density of electrons.  Thus, the eigenvalues of
the Kohn-Sham equations\cite{kohn-sham} were never supposed to represent the actual quasiparticle spectrum
of electrons.  Nonetheless, because the eigenvalues often, in fact, are a reasonably
good representation of the spectral properties measured in experiments,
this identification is commonly made in practice.
Hence, although everyone admits that this has no justification, most attempts to improve
BS theory are actually attempts to make corrections to the eigenvalue spectra to bring
it into agreement with various spectroscopies that probe the quasiparticle properties
of the materials, such as optical and photoemission spectra.  Even the metal versus insulator
problem involves this issue, since this distinction depends upon knowing the quasiparticle spectral
distribution as a function of energy.

From this very basic point of view, one could strongly question why
band-structure theory should be used for any spectral property,
since there is no formal justification
for such an application!  So, in this respect,
a correct starting point
for a HU-BS description should actually begin with an explanation of what spectral features
an LDA-BS description can be expected to accurately predict, and what many-body modifications need
to be made to improve this description.  
If LDA bands are to be used for the quasiparticle description of the non-$f$
electrons in this approach, 
it is important that this part of the theory should be placed on a firmer foundation.
As far as we know, this has never been done in any satisfactory way.

In order to examine this question, the best approach is probably
to consider the GW approximation\cite{hedin65,hedin-lundqvist}.  
Such a theory is developed in a Green's function formalism,
which is necessary in order to calculate spectral properties.  The one-shot
GW approximation can be written as an RPA-like correction
to any one-particle Hamiltonian, such as, for example, an LDA band-structure Hamiltonian.
GW theory has a formal derivation, and it is very clear what physics it includes and what it does not include.

In this type of approach, one could ignore the original derivation of LDA theory, and simply
treat the Kohn-Sham equations as an approximate one-electron representation of the electronic Hamiltonian.
The Green's function for this Hamiltonian can be used to calculate spectral properties, and,
if desired, to lowest-order in the screened Coulomb interaction, these results can then
be approximately corrected
to provide a better many-body theory of the electronic structure.  Framed in this way, one
can ask if better one-electron Hamiltonians than LDA would provide a better starting point
for spectral properties.  In fact, for example, ideas based on GW theory
for such an improved Hamiltonian
has been proposed by van Schilfgaarde, Kotani, and coworkers\cite{QSGW}.
Of course, better many-body corrections would then need to be added to any such one-electron
approach.

Historically, electronic-structure methods have forked into two paths.
The beginnings of this division were seen even 40 years ago\cite{hedin-lundqvist}:
``on the one hand, we have had the enormous wealth of \textit{energy band calculations}
which have had tremendous success in explaining the properties of specific solids,
but in which the connection with first principles is not always apparent.  On the
other hand, we have seen the spectacular progress of \textit{many-body theory}
applied to the solid state, which has given a number of new results, although
often of a rather general and formal nature, such as to provide the justification
and a formal basis for a one-electron theory."
In today's perspective, a very large effort has gone into 
improving the first-principles local-density approximations in order to
provide the best possible one-electron theory of electronic structure,
with the main focus on the accuracy of the total energy functional.
The advantage of this ``fork" is that such theories are usually first-principles (i.e.,
parameter free) and provide a detailed calculation of specific wave functions and
their spatial distribution with respect to the actual crystal structure of the material,
and also include the atomic number and core electrons of the relevant atoms.  It
is also usually possible to find the optimal atomic locations by energy minimizations.
The weakness of these types of theories is their poor treatment of the
many-body and quasiparticle aspects of the electron-electron interaction.
On the other hand, approaches in the second fork have attempted to focus on
this many-body character, albeit in the form of
simplified model Hamiltonians such as the Hubbard or Anderson Hamiltonians,
which can then be solved by a variety of sophisticated many-body techniques
involving various levels of approximation.

More recently, these two ``forks" have been merged into unified approaches, e.g.,
LDA+U (see, for example, Ref.~\cite{anisimov97} and references therein) or
dynamical mean-field theory, DMFT (see, for example,
Refs.~\cite{georges96,kotliar06,held07} and references therein), 
that are believed to include the best aspects
of both types of approaches.  These are the HU-BS methods mentioned above.
In order for theory to provide a proper guidance or context for various experiments
on different types of materials, it is essential to retain the details about the
types of atoms, their orbital character, and the atomic locations of the atoms
in the unit cell.  Otherwise, the calculations often become generic and less useful.
This is included in the BS part of the theory.
On the other hand, many materials clearly exhibit important many-body effects
that must be treated with more sophistication than LDA-BS methods.  This is treated
by many-body methods applied to the model Hubbard term in the theory.
Because these HU-BS theories have been ``built" on the BS Hamiltonian,
in the literature and at scientific conferences, one commonly
finds that many of these calculations
have been de facto considered as quasi first-principles methods.
It is the purpose of this paper to counter this prevailing assumption and
to provide a proper context for these new techniques.

As mentioned above, the types of electronic-structure calculations
that we will consider, in particular, are
all based upon adding an additional Hubbard-U term (or one of its variants)
to the band-structure Hamiltonian and then solving the resulting many-body
problem to some level of treatment.  When examining such approaches,
the critical question to ask is what such ``hybrid" approaches mean, or how
one should understand them.  How first principles are such approaches
and do they provide an adequate treatment of the electronic structure?
This type of discussion rarely occurs in the literature, but yet is crucial
if the field is to properly advance.

\section{What is a good band structure; what does a band structure measure?}

Since HU-BS methods are designed to correct band-structure calculations and to make them 
more realistic (i.e., to have better agreement with experiment), it is useful to
review what we may mean by a good band structure or what a band structure actually
measures.  In this regard, we can begin by reminding ourselves what goes into a band-structure
and what quantities result.  The input to a band structure are the atomic positions and
types of atoms (e.g., Cu or Si) within a unit cell.
The band-structure method then involves generating a one-electron Hamiltonian and calculating the electronic
wave functions and energy eigenvalues.  It also provides a total energy, and number density
(or charge density) and
spin density as a function of position.  From the energy eigenvalues the density of states
can be calculated.

The total energy as a function of unit cell dimensions and atomic positions is very useful since changes
in the total energy can provide forces on atoms that can be used in molecular dynamics programs, for
example, and can provide energy differences between different crystal structures.  
A good band structure could be defined in terms of how well it calculates this
total energy.  However, this
is not the main focus of this article.  We are more concerned with quasiparticles,
spectral properties, and the energy distribution of electrons.  These come from the energy eigenvalues
or dispersion relations (energy eigenvalues as a function of the k-vector in the Brillouin zone).

To answer the question of what a good band structure is and how we can experimentally confirm
such a band structure, one has to first ask first what are the fundamental intrinsic mathematical
formulations of the electronic structure of a solid and secondly how various experimental spectroscopies
are related to this formulation.  About 40 years ago, Hedin and Lundqvist wrote a very significant
review article\cite{hedin-lundqvist} that very clearly delineated the answer to these questions.

With respect to the first question, the answer surely must be that fundamental theoretical functions
that must be calculated are the one-particle and, more generally, $n$-particle Green's functions.  
The one-particle
Green's function, for example, provides information on the energy needed 
to add or remove one electron from the solid
as well as the energy dependent spectral density,
which can be written in terms of the imaginary part of the Green's function.  
The two-particle Green's function arises in a simple and direct calculation of
the total energy (although this can be reformulated in terms of the one-particle 
Green's function as well), as well as the dielectric 
and other response and correlation functions.  
Many of these are needed to evaluate neutron
scattering and magnetic response functions, for example, such as magnetic susceptibilities
or superconducting pairing.  
The value of the Green's function approach
is that it can incorporate simple approximations like one-electron approaches and yet can be
generalized to contain the full many-body physics of the electronic structure, also
including, for example, plasmons or other collective excitations.

When the one-electron band structure is put into a Green's function form, the results are very simple.  
The imaginary part of the Green's function is just
a sum over delta functions at the energies of the different eigenvalues.  Because the band-structure
is a one-electron theory each electron acts independently and excitations involve only
differences between the various energy eigenvalues with no correction effects.  Because the quasiparticle
spectra are just a series of delta functions, the lifetime of each quasiparticle is infinite
(there is no broadening of the spectral function by lifetime effects).  Also, because of the independent
particle approximation, there are no collective excitations.

For this reason, the predicted spectrum of the band-structure is simply a series of sharp quasiparticle
excitations (band energies as a function of $\mathbf{k}$).  
Corrections to this spectrum could come in two possible forms: (1) single-electron
corrections that would shift the energy eigenvalues as a function of $\mathbf{k}$, and (2) addition
of a frequency (and $\mathbf{k}$-dependent) self-energy that could also shift the effective quasiparticle
energies (through the real part of the self energy evaluated at the quasiparticle energy) as well
as provide quasiparticle lifetimes (through the imaginary part of the self energy evaluated at
the quasiparticle energy), and other-excited state effects such as satellite features at other
energies.  As we will see below, mean-field Hubbard model theories are examples of the first type of
correction, and dynamic many-body theoríes like DMFT the latter type.

To understand how ``good" this band structure is requires experimental verification of the spectral
properties or other ways of evaluating the Green's function that band-structure predicts.  This
is far from an easy task and in general involves correcting raw experimental data for a variety
of matrix elements, and other surface and experimental effects 
(for example, secondary electrons and experimental
resolution, etc.).  These issues are discussed later in this article (see Section~\ref{expt}).  
Here, it should only be noted
that many of these correction effects are often not carefully taken into account and our knowledge
of the ``experimental" spectral functions are probably not very good for most materials.

Finally, since HU-BS methods only correct the ``strongly correlated" orbitals and leave the other
(usually $s$, $p$, and some $d$) orbitals unchanged, the question of how well conventional
band-structure theory applies to the spectroscopic properties of these more extended orbitals is actually
very important and should be studied much more systematically than has been done up to now.

\section{The Hubbard term}

There are several important features about the HU-BS Hamiltonian
that must be emphasized.  First, the band-structure part of the Hamiltonian
identifies specific orbitals
and various hybridizations that provide a realistic description of the
underlying electronic structure and take into account
the correct underlying crystal structure.
In addition, such calculations are
first-principles and involve no adjustable parameters.
Secondly, the Hubbard term requires knowing the occupation numbers
of the ``correlated" $f$ orbitals (as mentioned above, we use the convention
that $f$ orbitals will be the correlated orbitals in this paper).

It should be pointed out that this ``hybrid" Hamiltonian has no derivation.
It is written down based on an intuitive understanding of the electronic structure.
The two terms are simply added together with no formal justification.  The
connection between the two terms comes through the $f$ occupation numbers in the
Hubbard term, which are assumed to be the same orbitals as the $f$ orbitals of
the underlying band-structure (the first term of the Hamiltonian).  Hence the
many-body treatment, which will only be applied to the second or Hubbard term involves
a projection of the Bloch states onto the $f$ orbitals.

The main assumption made by theories that involve adding a Hubbard
$U$ term is that band-structure calculations treat the Coulomb repulsion
between electrons at the mean-field level and that a more sophisticated
many-body treatment is necessary to handle strong electronic correlation
effects.  Thus the Hubbard $U$ term is reintroduced in a simplified way
so that a proper many-body treatment can be performed on this term.  To
avoid double counting, the mean-field evaluation of this term is subtracted
out in the belief that this removes the same amount of Coulomb repulsion
from the band-structure part of the Hamiltonian.  Hence, many-body effects
are then included at some level of sophistication while mean-field effects
are cancelled out.

It is important to consider whether these assumptions make any sense.
We believe that in fact they are seriously flawed.
For example, the Hubbard $U$ term is strongly
screened and appears nowhere in the original Hamiltonian,
which directly treats the explicit unscreened Coulomb repulsion.  In addition,
LDA calculations include a local exchange-correlation potential and hence
involve more than a mean-field (or Hartree) treatment of Coulomb repulsion.
A more straightforward approach would be to do a Hartree calculation of
the electronic structure and then add
an unscreened Hubbard $U$ term upon which to do the many-body treatment (including
screening).  This would be a disaster and such a theoretical approach would
lead to enormous errors in the electronic structure.

One useful way to assess the validity of adding this term to the BS Hamiltonian
is to take the local limit of this theory.  It is often asserted that HU-BS methods
become more exact as the correlated orbital becomes more localized.  An extreme
version of localization is to consider the isolated atom.  For example, it should be
possible to do both LDA+$U$ and DMFT for an isolated atom.  In addition, the constrained LDA
methods for calculating the effective $U$ should be very easy!

If this were to be done, the results would probably be very poor indeed.  Certainly the $U$
would not have the large screening of the solid and would most likely revert back to the
20--30 eV characteristic of the unscreened Coulomb integrals.  In addition, most of the
multiplet structure of the atom would be missing, unless it involved only direct $f$-$f$
multiplets and hence perhaps was specifically taken into account by the Hubbard-$U$ term.

This example is actually a very useful illustration of how dangerous it is to assume that
the HU-BS Hamiltonian is a good Hamiltonian to describe the overall electronic structure
of a system.  It directly demonstrates how strongly the HU-BS method
truncates the original exact Hamiltonian and therefore how severe this approximation is.
Is it possible to really assume that the screening of a solid can kill off so many aspects
of the electronic structure that such a simplified Hamiltonian as in the HU-BS method is
justified?  Thus, it shows very clearly how drastically we have reduced the actual complexity
of the electronic structure when applying HU-BS methods.  Obviously, one must take the
results from such theories with many misgivings.  In fact, as we argue elsewhere in this
review, it only makes sense to give up on the notion that these types of calculations are
first principles in any sense of this word, and that they can only reflect a convenient way to modify
the spectral weight of the band-structure predictions so as to better fit 
and interpret experimental data.

\section{Mean-Field Corrections}

Mean-field corrections to the original band-structure through the use of the Hubbard
term (i.e., LDA+$U$; see, for example, Ref.~\cite{anisimov97} and references therein)
provides an important illustration of how the addition of model Hamiltonian 
terms modifies and affects the original band structure.  These applications are
especially simple in that they do not change the one-electron character of the
Hamiltonian and hence can be solved simply and accurately.  In effect, they are
simply a slightly different band-structure than the LDA starting point.

In a Hubbard framework, these modifications are written in terms of the occupation operator
of specific orbitals (the ``strongly correlated" orbitals).  Hence they all have the same generic
form:
$$H_{MF} = \sum_{i m \sigma} V_{i m \sigma} \hat n_{i m \sigma} $$
where $V_{i m \sigma}$ is a function of the occupation numbers of the
$f$ orbitals on the same site $i$, and
$\hat n_{i m \sigma}$ is the number operator for the $f$ orbital $i m \sigma$.  
In the mean-field approximation this 
is usually a linear function of the occupation numbers 
$$ V_{i m \sigma} = V_{i m \sigma}^0 +
\sum_{m' \sigma'} U_{i m \sigma ,i m' \sigma'} n_{i m' \sigma'}, $$
where $V_{i m \sigma}^0$ and $U_{i m \sigma ,i m' \sigma'}$ are numerical constants, and
$n_{i m' \sigma'}$ are the occupation numbers of the $f$ orbitals (which have
to be solved self-consistently in the course of the calculation).  
This approach can, of course, also be generalized for nearest-neighbor
or more distant Coulomb-like interactions.

The first point to note about these relationships is that they depend on the number operator
of the correlated orbitals.  Hence they are orbital-dependent interactions and require a projection
of the electronic states onto
the number density on these orbitals in order to specify the interaction.  Thus they depend
specifically on the basis set that is used.  Intuitively, they are meant to be intra-atomic
corrections, so that one prefers that these orbitals look as atomic-like as possible.  There
are actually two different choices that can be made in this regard.  Since many BS methods
involve muffin-tin basis sets that have specific numerical wave functions for each type of orbital
angular momentum, one could project these occupation numbers onto an occupation number for
only these parts of the wavefunctions.  As the wavefunctions in a solid extend both into the
interstitial region and other atomic spheres, such occupation numbers would always be less than
one when projected onto the radial wave function of any specific sphere.  Alternatively, one can view
these atom-centered basis functions to be Wannier functions centered on each site, and to use
maximally localized Wannier functions so that the portion of each Wannier function has as much atomic-like
character as possible on the relevant atomic center.  Each of these choices has some drawback.  The first
choice is somewhat ill-determined since it involves the way the wavefunctions are normalized, and the
second choice puts parts of the Wannier function into the interstitial region and onto the $s$, $p$,
and $d$ orbitals of other atoms and hence loses some of the intra-atomic character that is being
corrected for.  Both choices depend on the size of the muffin-tin radii and how much of the
wavefunctions are localized within a given sphere.  Either choice is only likely to be somewhat
satisfactory for very localized or atomic-like wavefunctions and to become less well defined as
the wave functions become more diffuse.  In practice, there is some interplay between the value of the projection
used and the parameters of the model Hamiltonian term that is used.  For example, if the method
chosen for the projections leads to smaller occupation numbers, one can correct for this by
increasing the values used for the mean-field (i.e., the Hubbard-$U$) parameters.

The first correction factor ($V_{i m \sigma}^0$) is usually described as a ``double counting"
term and simply shifts the bare energy level of that specific $f$ orbital up or down in energy. It can be
used to precisely place the energy of any given orbital wherever it
needs to be in order to achieve good agreement with experiment.  If it is independent of the
z-projection of the orbital ($m$), it simply shifts all the $f$ orbitals up or down.  An alternative
way of viewing this correction is as a way of modifying the occupation number of any orbital.
By shifting their energy up or down, one controls how much of the orbital is occupied.  For example, this
type of interaction is identical to that which is used as the Lagrange parameter in constrained
calculations of the Hubbard $U$.

In the literature, different choices are made for this first correction factor for
different ``flavors" of mean-field theories.  Since the ``double
counting" is actually an illusion (one is actually not adding and substracting terms from
the original starting Hamiltonian), the only satisfactory way to choose which method
one wants to employ (or perhaps to add even a different constant shift of the orbitals)
is to compare with experiment.  Otherwise there is no fundamental physics argument to
choose one method in preference to the other.

The second term in mean-field theories (involving $U_{i m \sigma ,i m' \sigma'}$) is an orbital polarization
term.  It causes different $f$ orbitals to shift in terms of their relative energies depending on
the specific occupations of each orbital and on the values of the coefficients $U$ that are chosen
(especially if they are positive or negative).  Given this functional form, any polarization that
is desired in order to fit experiment can be forced upon the band-structure
solution if the proper coefficients are chosen to do this.

In addition to these effects, the orbital dependence of these Hubbard terms also makes it possible to
include non-local exchange effects, since orbital-dependent interactions
can be used to represent a non-local function.
For example, in pseudopotential theories an $l$-dependent potential is often added to represent
the non-local character of the pseudopotentials.
Since the starting LDA potentials use
a local exchange potential, the Hubbard terms can be a way
of correcting the LDA band structure for non-local exchange.  
In the quasi-particle self-consistent method screened non-local exchange interactions
coming from the GW approximation are included as orbital-dependent potentials to
correct the one-electron band structure\cite{QSGW}.
Similarly, 
it has long been known that the LDA-BS method suffers from
self-interaction errors, and the LDA+U may be viewed as a method to remove
self-interactions. In Hartree-Fock theory, for example, the self-Coulomb and
self-exchange interactions exactly cancel, but once the exchange interaction
is described in terms of a local potential, this cancellation is no longer
exact. The screening of the Hubbard U parameters may then be justified by
the fact that part of the self-exchange is being removed by the local
potential.

\section{Dynamical Corrections}

The mean-field treatments considered in the previous sections are basically different variations
on band-structure calculations.  All are one-electron theories and, at best, simply modify
the disperson relations of the bands (energy versus $\mathbf{k}$).  However, as was well
explained in the early classic paper by Hedin and Lundqvist\cite{hedin-lundqvist}, the
exact Green's function for the electronic structure contains a significant frequency dependence
in its self energy.  Since a band-structure calculation is a static approximation for
the electronic structure, it has no frequency dependence, and completely misses this structure.
Hence, if one is going to correct band-structure theory in order to provide
a more realistic electronic structure, it is essential to consider
how to incorporate self-energy effects.

The GW approximation, as its name suggests, automatically generates a self-energy that
is proportional to a Green's function times a screened Coulomb energy.  Although this self energy
is the lowest order term in an expansion in the screened Coulomb energy, it still incorporates
some important features that more sophisticated treatments will need to include.  For example, for
weakly correlated systems it maintains the quasiparticle structure inherent in the band structure.
Hence the spectral function often has a strong peak at the quasiparticle energy.  The energy
of this peak can be considered to represent the corrected band-dispersion relations and the width provides
a lifetime for the quasiparticle.  It can also correct the size of band-gaps in semiconductors
and open gaps in systems that otherwise would be metallic within LDA band-structure theory (although
sometimes this requires using LDA+U or other theories to first create a gapped electronic structure
as the starting point for a one-shot GW calculation).  Because it incorporates a non-local screened exchange
term, it can also provide the type of corrections that traditionally have come from Hartree-Fock-like theories,
for example, such as are often added by LDA+U approaches.  Hence it can account for some of the modifications
discussed in the previous section on mean-field approaches.

Although simpler many-body approaches can be incorporated into HU-BS
approaches\cite{steiner91,steiner92,steiner94},
the state-of-the-art methods are now almost exclusively DMFT.
This involves a non-perturbative many-body solution of the Hubbard term
that is performed by mapping the original problem onto a single-impurity
Anderson model (SIAM) and solving the SIAM as exactly as possible.
It requires a projection onto the strongly correlated $f$ orbitals.
The method produces a self-energy coming from the $f$ orbitals only.
These can generate satellite spectra (lower Hubbard bands) as well as
Kondo-like peaks at the Fermi energy and large specific heat enhancements.
They will also provide an electronic lifetime for states that have a significant
$f$ character.  However, these lifetimes only come from the $f$-electron self energy, while
the other $s$, $p$, and $d$ electrons have no self-energy or lifetimes and
are the original starting band-structure dispersions.
Such theories are useful if electronic correlations dominate
the ``interesting" parts of the electronic structure.
The Hubbard-$U$ parameter must either be estimated from constrained
Hubbard-$U$ calculations or fit to experiment.  Although
much success has been claimed for these types of theories, the
experimental verification is often qualitative.
The critical assumption of the single-site DMFT is that the self energy
of the correlated electron states is $\mathbf{k}$ independent.

One very serious issue with DMFT approaches is the ``solver" for the
SIAM equation in the theory.  At the present time, many different
solvers are used.  
Most SIAM solvers, whether from iterative perturbation theory or non-crossing
approaches, etc., have large uncertainties in the correctness of the
many-body solutions they provide.
Only the quantum Monte Carlo and numerical
renormalization group solvers are exact.
However, even for these, 
despite new algorithmic advances such as
continuous time quantum Monte Carlo techniques,
there is considerable uncertainty about
the quality of their results. For example, the
quantum Monte Carlo methods requires an analytic
continuation from the imaginary frequencies that are calculated by
the method to the real frequencies needed for physical properties.
Even with new and sophisticated techniques for accomplishing this analytic continuation such as
those involving maximum entropy, the real frequency results are
very sensitive to small changes in the imaginary frequency results
leading to concerns of large errors.  Also, the Monte Carlo solvers
are most accurate at high temperatures and become increasingly 
untrustworthy at low temperatures where most of the interesting correlation
physics lies.  Overall, from the point of view of the SIAM solvers,
at this time one has to strongly question the accuracy of most DMFT calculations.
In particular, it is clear that different solvers will give different results,
as emphasized at the beginning of Sec.~5 of Held's DMFT review.\cite{held07}
  
In general many-body theories must be added to band-structure
methods if the correct electronic structure is to be produced.
However, self-energy effects need to be generalized to correct
the non-correlated orbitals as well as the correlated orbitals.
All of the quasiparticles (except those exactly at the Fermi surface)
have finite lifetimes and are likely to require corrections to
their dispersion relations relative to the LDA starting point.
In addition, plasmon, lower Hubbard band, and other non-quasi-particle
features will in general be present in the electronic structure.  Such
effects are not included in the LDA band structures.

\section{Is the HU-BS approach a real electronic structure method?}

At this point, based on the previous discussion, it is useful to summarize
our review of the content of the electronic structure implicit in the
HU-BS methods.  The most striking comment that can be made on this method is the starting
Hamiltonian itself, Eq.~(\ref{HU-BS-eqn}).  Compared with the exact
Hamiltonian, Eq.~(\ref{exact-eqn}), it is clear that such a drastic simplification has
been made that the credibility of the HU-BS Hamiltonian cannot be taken at face value
but must be carefully assessed.  Exactly what has been done?

From the form of Eq.~(\ref{HU-BS-eqn}) and the fact that the Hubbard term is
a model term whereas the first term is an attempt at a first-principles description
of the electronic structure, it is reasonable to interpret this Hamiltonian
in terms of its most fundamental part, the band-structure Hamiltonian, and a correction
term, the Hubbard term.  In addition it is commonly assumed that the double-counting
term is just the same term but treated in the same mean-field way as
the local-density approximation, and thus
one is essentially adding or subtracting
the same effect in order to do a more exact treatment of 
the most difficult part of the physics.  However, this is not really credible.  
Neither the Hubbard term nor the double-counting term exist in the original Hamiltonian.
They simply represent an ``ansatz" that has been inserted by hand.  Thus, they can
only be sensibly understood as a ``correction" to the
band-structure Hamiltonian.  These terms are a means by which to include
additional many-body physics that was left out when the rather drastic approximations needed
to formulate the LDA Hamiltonian were made.  Hence they make it possible to build in new
features such as satellite peaks and to adjust the quasiparticle spectra of the
band structure.

To approach Hubbard $U$ theories in this spirit provides new flexibility
and should make it easier to resolve certain controversies that
often arise, such as which LDA+U theory is the best approximation.
For example, once it is realized that double-counting is not an issue, one can focus
more on what electronic-structure effects
have been left out of the band-structure approach and what ``model" terms
could best correct for these effects with a better many-body treatment.
In fact, one could question, for example, whether other expressions
that are different
from the Hubbard $U$ term would lead to better corrections or whether
one should instead add corrections to the self-energy of the electronic
Green's function instead of adding additional terms to the Hamiltonian.

An important consideration is whether the HU-BS approach can actually work.
How do we know what physics is left out, and why do we believe that the
model term is the right correction factor?  Finally, can we actually do
the many-body physics in a sufficiently correct way to believe that
we have significantly improved our understanding of the electronic structure?
A bad treatment might actually lead us to a worse description.  Also,
an important aspect of this approach is that we need to include parameters in
the theory in order to mask our ignorance of the real many-body microscopic theory
that we are at present unable to successfully attack.  What are the implications
of being forced into a parameterized approach?

Before delving into such matters, however, it is useful to examine more closely the
Hubbard-$U$ term again.  If $U$ is treated as a matrix and allowed to depend on the $m$
projection of the $f$ orbitals, this term is exactly of the same form as the original 
Coulomb integrals for a fixed basis of $f$ orbitals.  In this sense it has the same
physics as the Coulomb term for a fixed (or frozen) atomic basis, although the basis functions are
limited to one type ($f$ orbitals only) and these functions are extremely limited in scope
(a minimal $f$ basis).  If used for an atomic calculation, such a limited basis would give
very poor results for treating the electron-electron Coulomb terms.  So, why should we expect
an accurate treatment in the solid?  We believe that, in fact, this term does not provide an
accurate treatment.  The $U$ matrix that is used in the HU-BS approximation is heavily screened.
What is actually going on is that the many-body treatment of the Hubbard-$U$ term is being
used on something that looks like the original Coulomb term.  Hence, the form of the results
(the types of peaks and excitations in the Green's function) has a frequency dependence
and quasiparticle spectra similar to what a real Coulomb term would generate.  By scaling
down the Hubbard $U$ one reduces the strength of this effect while retaining the same functional
form (freqency or spectral dependence).  Thus, if the original electronic structure is missing
peaks or features, this is a way of reintroducing them.  At the same time one has a tuning parameter
that can be used to fit the peaks in an experimental spectra.  Thus, such a theory provides realistic
spectra that can be fit to experiment, and 
with which to correct the LDA band structure for these types of missing
features.  Because it is not the electronic structure calculation of any actual electronic-structure
Hamiltonian but of a model or pseudo-Hamiltonian, the accuracy of the results really doesn't matter.
As long as the right types of peaks or other features that are seen in experiment are present in
the many-body results, the Hubbard-$U$ parameter can be scaled up or down 
to fit the experimental peaks or features.
Essentially, the HU-BS approach is just a model solution of a Coulomb-like term, with the final
results scaled and then mixed with some LDA band structure.

In practice, as discussed above,
mean-field treatments of the Hubbard-$U$ term are used to add in Hartree-Fock
like atomistic structure into the one-particle spectra.  These can essentially orbitally
polarize the correlated $f$ shell of electrons.  They can also account for SIC-like corrections.
For dynamical theories like DMFT, two effects are commonly introduced: (1) an additional peak
in the photoemission (the lower Hubbard-$U$ band), and a narrowing of the correlated quasi-particle
bands.  These are all that are required to fit the experimental data.  Besides the Hubbard-$U$ parameter
itself, additional parameters such as the Hubbard-$J$, etc, can be added if the single $U$ parameter
is too crude to fit the experimental data.  Hence, since there are plenty of available parameters
and such limited data set with which to fit to, the HU-BS approach is almost certain to be
in good agreement with experiment.

This argument could be turned upside down, of course.  Is there any experimental data
that show the essential correctness of the HU-BS approach other than being
a simple fitting procedure? 
We have been unable to find any such examples, which
leads us to the conclusion that HU-BS approaches are simply ways to add in some crude
many-body effects that are left out of the original band-structure calculations.
Similar questions could be framed in another way.  For example, 
are there any surprises from these types of calculations that
could not have been guessed from the model calculations
alone?
How much physics does the band-structure piece of the Hamiltonian add
that is not included in the Hubbard-U term?
What new physics has really come from the merging of these two Hamiltonians?
Is there any rigorous confirmation of DMFT or any other HU-BS approach that
goes beyond a fit to some experimental data?

The ideal approach would, of course, be to start from the exact microscopic theory of electronic
structure and then to make various approximations that then lead to different levels
of sophistication in the solution.  This is similar to the line of theories starting with
Hartree and Hartree-Fock solutions, through various flavors of local-density approximations,
and up through GW-type theories.  However, at this point our abilities to calculate true
many-body effects from first principles appears to have hit a dead end, in the sense
that it is unclear how to go further with a tractable theory.

This is, in fact, the driving force for developing Hubbard-$U$-like approaches.  Modern
many-body theory has heavily focused on solving simplified electronic-structure Hamiltonians
based on a simple nearest-neighbor tight-binding treatment of the Hubbard Hamiltonian, often based
on a single orbital per unit cell.  By simplifying the electronic structure, it was possible
to focus on the complex mathematical manipulations that are necessary to treat the many-body
aspects of the theory.  The price that was paid for this approach was to lose the connection
to the specific types of orbitals, atoms, and their geometries possessed by real materials.
Hence one ended up with ``spherical cow" approximations to the electronic structure of materials
that could not well describe Fermi surface or photoemission details of materials of interest.
To include these material-dependent properties the LDA band-structure was then added back into the various approaches, with the treatment of the Hubbard $U$ term projected onto the most localized
or strongly correlated orbitals.  Since the model calculations were viewed as simplifications
of the real electronic structure, this lead to the conclusion that one had to add and subtract
terms from the band-structure Hamiltonian, leading to the notion of double-counting, etc.

We believe that the correct many-body treatment of the microscopic electronic Hamiltonian
is still too difficult for current levels of theory.  Hence some simplifications of the
many-body effects will require the introduction of model terms or expression that parameterize
corrections of the first-principle theory.  What these additional terms do in practice is
to push spectral weight of the electronic structure away from that calculated by the original
band-structure theory.  For example, in some materials, remnants of the original atomic structure
show up as satellite features (often described as lower Hubbard bands) below the conventional valence
band structure or as additional peaks in the density of states at the Fermi energy (often described
in terms of Kondo effects in many theories).  These effects cannot naturally arise in the one-electron-type
approach of band-structure theory.  Since they cannot be calculated from first-principles,
one has to add in parameters to the theory to force the electronic-structure theory to agree
with available experimental data.  One can then question how best to correct the original
band-structure theory to force this agreement, and what understanding such a theory provides
about real electronic structure that an exact theory would predict.  There are also questions about the
robustness of such corrections.  For example, is each correction materials specific, or can trends
be determined for classes of materials that continuously tune these parameters.  Also, is there
enough physics in the ``correction terms" to allow one to understand the correct mechanisms controlling
the functionality and many-body properties described by such theories?

\section{Experimental Verification}
\label{expt}

The key to making progress is good experimental data.  
Since we cannot trust the current level of theory to
accurately predict materials properties, especially when part of the electronic
structure depends upon unknown parameters, experimental data is necessary to guide theory.

The chief obstacle with respect to experimental data is that most experiments do not directly
measure the fundamental mathematical properties of the electronic structure, viz., the
various Green's functions and spectral densities of states.  If we need completely
different electronic-structure theories for each specific type of spectroscopy, we will
only attain many random bits of information that do not form a coherent whole.  To be useful
there has to be a common ground where all the experimental data converge.  This common ground
has to be a fundamental property of the electronic structure and it must be amenable to theoretical
techniques.  Hence it makes sense to focus on spectral and other fundamental properties of the electronic
structure.

From this point of view, one has to ask what each type of spectroscopy measures and how each one
can shed light on the various Green's functions or their spectral representations.  This will
depend upon a very clear theoretical understanding of all of the physical processes that
are involved in each spectroscopy and how to account for these in order to pull out of the raw data
the fundamental information about spectral functions.  At the present time, very little emphasis
has been placed on this.  Most interpretations of experimental data are very simple minded and have
not changed much in the last 30 years.  Spectrometers and the physical and electronic equipment
used in the various techniques has undergone enormous improvements, but this is not the case
for the fundamental theory needed to interpret the data.  
This is a situation that desperately cries out for improvement.
The best way to advance our understanding of strongly correlated electron materials, for example, is to
improve our understanding of what each spectroscopy accurately measures about their properties.  This can
only be achieved if we have a proper understanding of the fundamental physics of each spectroscopic
method.  

While one could discuss many different types of spectroscopy here, probably the most
heavily used spectroscopies that most directly measure electronic spectral densities
involve interactions of photons with matter, such as optical spectroscopy and photoelectron
spectroscopy.  In each case, the fundamental process involves the absorption of a photon
by exciting an electron from an occupied state of the solid to an unoccupied state.
It is useful to do at least a brief exploratory discussion of how these types of experiments can
be related to the fundamental electronic stucture.  Here we will only discuss photoelectron
spectoscopy as a prototype for the types of discussion that need to be more generally employed.

The simplest theories of photoemission (see, for example,
the recent review paper on the cuprate superconductors, Ref.~\cite{damascelli03},
that cites many earlier review papers, as well as the standard book
on the subject, Ref.~\cite{hufner}) use the three-step model developed in the early days by Spicer
and coworkers.  This treats the photoemission process as involving: (1) an electronic excitation of
the system by a photoelectron, (2) the transport of the photoelectron to the surface,
and (3) the escape of the photoelectron through the surface to the vacuum where it is detected.
Even this very simplified model already hints at how complicated photoelectron
spectroscopy really is.  Not only excitation processes need to be described, but electron transport
and detection as well.  Also, the surface clearly must play an important role.

If we just focus on the first process, the excitation of the electrons by the photon, even
this is not simply related to the spectral function of the desired one-particle Green's function.
Clearly, this must involve the transition from an occupied electronic level to an unoccupied.
In a band picture, one would calculate this from the Golden Rule, and this would involve
occupied electrons of a given band index and $\mathbf{k}$ being excited to a higher lying band with
some matrix element squared (which would also, of course, have selection rules).  This involves
a convolution of occupied and excited states.  From this, how is the
spectral function $A(\mathbf{k},\omega)$ to be determined?

In practice, the cuprate review article\cite{damascelli03} (see their Eq.~(12))
recommends using the sudden approximation and the formula for the observed
electron intensity:
$$ I(\mathbf{k},\omega) = I_0(\mathbf{k},\nu,\mathbf{A}) f(\omega) A(\mathbf{k},\omega)$$
where $\mathbf{k}=\mathbf{k}_{\parallel}$ is the in-plane momentum,
$\omega$ is the electron energy with respect to the
Fermi level, $I_0(\mathbf{k},\nu,\mathbf{A})$ is proportional to a
squared one-electron matrix element and therefore
depends on the electron momentum as well as the energy and polarization of the incoming
photon, and $f(\omega)$ is the Fermi function.  
The function $A(\mathbf{k},\omega)$ is the electron spectral
function.

With even this additional level of simplification, analysis is still not completely simple.
In most experimental papers, it appears as if the angle-resolved experiments
simply track peaks in the observed
energy distribution curves as a function of angle and energy.
These peak energies are then plotted
relative to an estimated Fermi energy to produce an ``experimental" band structure.
However, what about the matrix elements $I_0(\mathbf{k},\nu,\mathbf{A})$.
If these are strongly k or energy dependent, they could
certainly drastically distort apparent band positions.
One also has to question how valid the sudden approximation is
(see, for example, the classic discussion in Ref.\ \cite{hedin99}).  
The chief argument for such a
simplified analysis of photoemission is that the results appear
somewhat similar to one-electron band-structure calculations!

In the above formula, one has to question what happened to the unoccupied
states that the photoelectron is excited to?  H\"ufner's book suggests
using free-electron band-structure expressions for accounting for this quantity,
which would involve a more complex analysis than given by the above formula.
However, electronic band-structure calculations suggest that there is significant
band-structure effects that strongly distort even fairly high energy unoccupied electrons
away from their free-electron energies.  This is likely to be the case for the relatively
low energies used in the UV photoelectron range, which has the highest precision.  What
are these corrections and how much do they change the determination of $A(\mathbf{k},\omega)$?

Finally, one should consider the effects of the surface and a large number of other
physical processes such as secondary electrons that complicate the experiment
(see, for example, the H\"ufner book).
Even 35 years ago, this complexity was recognized as important for understanding the comparison
between band theory and photoemission experiments\cite{christensen74,feurbacher74}.  
However, today, most of this complexity
seems to have been swept under the rug, with simple peak evaluations trusted as reliable
estimates of the spectral functions!
It is certainly incumbent upon the experimentalists to correct their data as carefully
as possible in order to provide the best experimentally determined spectral function as possible.

To us, one of the most problematical aspects of photoemission is its high surface sensitive.
Besides possible effects in shifting the peak positions in which quasiparticle energies
are based or on the appearance of new surface electronic-structure peaks,
another example of problems with surface sensitive spectroscopies is the danger of artifical narrowing
of strongly correlated electron bands.  Tight-binding theory suggests that the order of magnitude
of the band width is proportional to the number of near neighbors times the nearest-neighbor
hopping matrix element (see, for example, the analysis in Ref.\ \cite{christensen74}).  
At a surface, there are fewer nearest neighbors and bands should narrow.
This is actually observed in LDA-like calculations of surfaces.  See, for example,
Ref.\ \cite{eriksson}.  Correlation effects may artificially enhance such effects, leading
to a significant narrowing of bands that is purely a surface effect.  In the HU-BS methods, band
narrowing can be caused by increasing the value of the Hubbard $U$.  If this is fit to photoemission
that is really measuring the surface band width, significant errors may be introduced and misleading
conclusions drawn.  Perhaps many strongly correlated materials are far less
correlated than they appear, and the band narrowing observed in experiment is
just a measure of enhanced ``surface" electronic structure?
It is unknown how surfaces may modify satellite, Kondo, and other many-body features.
Again, it is possible that they greatly enhance such effects.

While it is clear that these types of experiments desperately need a good
theoretical underpinning to aid in the interpretation of the data,
on the other hand, the very fact that such experiments are
surface sensitive makes it very difficult to develop the precise theory needed to interpret them.
Surfaces introduce changes that depend on how they are prepared and are a much less intrinsic property
of a material than bulk electronic structure.  Often the presence of oxygen, hydrogen, or other impurities can
significantly modify the nature of the surface.  In addition, there is the possibility of preferential
segregation of bulk impurities to surfaces, particularly if some heat treatment or annealing has been
performed.  For a good theory to be developed, it is necessary to have a very precise knowledge of
all of the atomic positions and types of atoms at a surface, before attempting to account for 
the excitation process of the photon, and the transport of the resulting photoelectron to the surface,
emission through the surface,
and its collection.  This involves very complex physics, and is quite difficult.  However, until we
have a better understanding of the theory of photoemission, and how the nature of these types of
measurements affect the resulting electronic properties that are measured, it will always be somewhat
dangerous to rely upon such experimental data to tune the parameters of a strongly correlated material.

In addition to surface sensitivity, lifetime effects can also be problematic, and may limit precise measurements
to energy regions around the Fermi energy.  Usually the lifetime of an occupied electron state
increases rapidly as
its energy moves farther below the Fermi energy.  In photoemission, this rapidly washes out the experimentally
determined dispersion relations and it becomes difficult to know what the quasiparticle energies are deep
(or even moderately) below the Fermi energy.  Since the electronic lifetimes are an intrinisic bulk effect, 
this effect cannot be reduced in any type of spectroscopy.  This can, for example, make it difficult
to measure shifts between the bottom of the $s$ band and the position of the bottom of the $f$ band
in actinides, which might be useful to know if one wants to understand how nonlocal exchange potentials
shift localized electronic states relative to delocalized.

Viewed more broadly, excitation spectra can have also additional effects that are unrelated to ground-state electronic
structure, making it difficult to know what the intrinsic electronic structure is in the absence
of the specific experimental probe used to measure the electronic structure.  A well known example
of this is exciton effects in semiconductors.  In an optical probe, the incoming photon excites an occupied
electron to an unoccupied state, creating an electron-hole pair.  The electron and hole repeatedly
scatter off of each other (this is usually calculated by a Bethe-Salpeter equation) and the resulting
excitation lies in the energy gap of the semiconductor.  If not accounted for, this would erroneously lead
to a conclusion that the intrinsic band gap is smaller than it actually is.  Other issues are explicit
excitation processes that are different from ground-state electronic structure, such as shake-up,
shake-off, and other multiplet or other intra-atomic processes involving electronic excitations
that occur nearly simultaneoulsy with the one-elecron process of interest.

Given these experimental difficulties, one can question how well we know the experimental electronic
structure, and whether the many-body corrections that we are including by fitting to such
data is really correct.  There is thus a real need to develop a better theoretical underpinning
for the various experimental techniques that are being used so that we can more reliably
interpret such data.

\section{Summary}

HU-BS methods involve adding a Hubbard model term to an LDA band-structure Hamiltonian.
The Hubbard model term is a static Coulomb interaction for frozen orbitals with matrix
elements that are scaled to fit experiment.  A double-counting term is actually
just a way of preventing the average energy of the correlated orbitals from being pushed too high
in energy and should be considered just another parameter of the theory that fits
the correct average occupation of the correlated orbitals.  The LDA band-structure
is a model for the non-correlated orbitals.  It replaces the small number of nearest-neighbor
hopping terms of traditional model Hamiltonians with the full complexity of all of the
relevant orbitals of the various atoms in the solid.  However, it suffers from the defect
that LDA does not correctly treat the spectral properties of these orbitals.  In particular,
non-local exchange and self-interaction correction effects are improperly treated.  The
accuracy of the HU-BS methods cannot be determined very well, because it is difficult
to correct any current spectroscopy sufficiently to accurately measure the intrinsic
spectral functions of the electrons in the solid.  In practice, the HU-BS methods add
lower Hubbard band peaks and narrow the band-width of the correlated states.  The linear
term can also be used to introduce Hartree-Fock like structure to open band gaps and orbitally polarize
the electrons.  Because these methods use parameters, they are fits to the experimentally
observed spectra (whether these are an accurate measurement of the actual spectral functions
or not) and are not first-principles methods.  They should be viewed as simply more elaborate
model calculations to include more orbitals than traditional Hubbard models, which often only
have one or a very small number of orbitals.
Because the LDA term takes care of the non-correlated orbital interactions, the number of fitting parameters
of a traditional Hubbard model is reduced for these extra orbitals at the price of the loss
in accuracy entailed by the LDA method.

Progress in the future has to involve two aspects.  The first is better first-principles
starting points that incorporate more and more of the correct physics.  The better these are
and the more physics they incorporate, the fewer the corrections that need to be made
to compare with experiments.  Secondly, better solutions to a variety of strongly interacting
models are needed.  What does the frequency dependence of the exact self-energy
for these various models look like?  Will they show any surprises, such as additional
features in their frequency dependence?
If, for example, the exact theory simply shifts the lower Hubbard
sattelite away from the position of a less accurate theory, this could be corrected for by simply
modifying the strength of the Hubbard $U$ parameter.
What is particularly important here, and which has not been carefully examined in the past,
is the generic functional forms that the many-body solutions involve. 
For example, the Hubbard Hamiltonian
usually causes lower and upper Hubbard-band satellite features.  What is the functional form of the
frequency response of the self energy that causes such satellite structure to appear in the
spectral response of the many-body system?  Could one parameterize this in such a way as to compare
different many-body solutions and understand how to include model frequency dependent functional
forms in self-energy corrections to the starting band-structure solutions?  Could one model these in
a way similar to what is done in Fermi liquid theory?  For example, if one knows that some quadratic or perhaps
exponential frequency dependence must show up in the exact theory, could one simply parameterize
this response to correct the band structure?  

Finally, much more work must go into the interpretation of various spectroscopies if these are to
be accurately related to the bulk spectral functions predicted by theory.  Without good experimental
bulk spectral functions, it is impossible to tell how good or poor our current models of electronic
structure are.  Without a true first-principles method, the parameters of the HU-BS models
must be fit to experiment.  The resulting electronic structure 
will heavily depend on the quality of the experimental
data that is fit to.




  

\ack
This work was carried out under the auspices of the
National Nuclear Security Administration of the U.S.
Department of Energy at Los Alamos National Laboratory
under Contract No. DE-AC52-06NA25396,
and the Center for Theoretical Natural Science at Aarhus University.

\section*{References}


\bibliographystyle{unsrt}

\bibliography{hubs}


\end{document}